\documentclass{elsart41}
\usepackage{graphics}
\usepackage{graphicx}
\usepackage{amssymb}

\newcommand{\tal}{TlCuCl$_3$}

\begin{document}

\begin{frontmatter}



\title{Uniaxial pressure dependencies of the phase boundary of \tal}


\author[Colonia]{N. Johannsen},
\author[oos]{A. Oosawa},
\author[tan]{H. Tanaka},
\author[Mos]{A. Vasiliev}, and
\author[Colonia]{T. Lorenz\corauthref{Lorenz}}
\ead{lorenz@ph2.uni-koeln.de}

\address[Colonia]{II.\,Physikalisches Institut, Universit\"{a}t zu
K\"{o}ln,Z\"{u}lpicher Str. 77, 50937 K\"{o}ln, Germany}
\address[oos]{Advanced Science Research Center, Japan Atomic Energy Research
Institute, Tokai, Ibaraki 319-1195, Japan}
\address[tan]{Department of Physics, Tokyo Institute of Technology, Oh-okayama,
Meguro-ku, Tokyo 152-8551, Japan}
\address[Mos]{Department of Low Temperature Physics, Moscow State
University, Moscow 119992, Russia}

\corauth[Lorenz]{Corresponding author:}

\begin{abstract}
We present a thermal expansion and magnetostriction study of \tal
, which shows a magnetic-field induced transition from a spin gap
phase to a N\'{e}el ordered phase. Using Ehrenfest relations we
derive huge and strongly anisotropic uniaxial pressure
dependencies of the respective phase boundary, e.g.\ the
transition field changes by about $\pm 185$~\%/GPa depending on
the direction of uniaxial pressure.
\end{abstract}

\end{frontmatter}


The magnetic subsystem of \tal\ may be viewed as a set of
three-dimensionally coupled spin dimers with a singlet ground
state. Due to the rather strong interdimer couplings the first
excited triplet state has a pronounced dispersion and as a
consequence the minimum spin gap $\Delta \simeq 8$\,K is
significantly smaller than the intradimer coupling $J\simeq
64$\,K.\cite{Nikuni2000matsumoto2002ruegg03} Thus a magnetic
field of about 6\,T is sufficient to close $\Delta$. Above 6\,T a
field-induced N\'{e}el order with transverse staggered magnetization
is observed,\cite{tanaka2001Oosawa2002d} which has been described
by a Bose-Einstein condensation of
magnons.\cite{Nikuni2000matsumoto2002ruegg03} The ground state of
\tal\ is very sensitive to pressure, hydrostatic pressure of
order 0.5\,GPa is sufficient to induce the N\'{e}el order already in
zero magnetic field.\cite{Oosawa2003atanaka2003a} In order to
obtain more information about this drastic influence of pressure
we performed high-resolution measurements of the thermal expansion
$\alpha_i=\partial \ln L_i/\partial T$ and the magnetostriction
$\epsilon_i = (L_i(H)-L_i(0))/L_i(0)$ and derive the uniaxial
pressure dependencies of the transition temperatures and fields.

\begin{figure}[b]
\begin{center}
\includegraphics[width=0.31\textwidth]{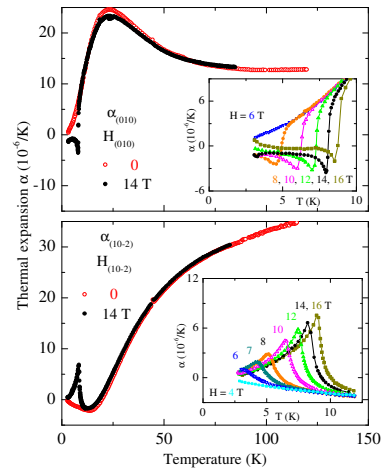}
\end{center}
\caption{Thermal expansion $\alpha_i$ perpendicular to the $(010)$
and the $(10\overline{2})$ cleavage planes of \tal\ for different
magnetic fields ($H||\alpha_i$).
} \label{fig1}
\end{figure}

\begin{figure}[t]
\begin{center}
\includegraphics[width=0.37\textwidth]{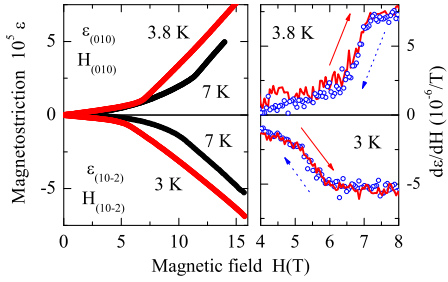}
\end{center}
\caption{Left: Longitudinal magnetostriction $\epsilon_i$
($H||\epsilon_i$) perpendicular to the $(010)$ and the
$(10\overline{2})$ cleavage planes of \tal\ at different
temperatures. Right: Field derivatives $\partial \epsilon_i
/\partial H$ from the measurements with increasing (---) and
decreasing ($\circ$) field.} \label{fig2}
\end{figure}

Fig.\,\ref{fig1} shows $\alpha_i$ perpendicular to the $(010)$
and $(10\overline{2})$ cleavage planes of \tal\ measured for
different magnetic fields. In zero field $\alpha_{010}$
($\alpha_{10\overline{2}}$) shows a broad maximum (minimum) around
18\,K. This anomalous temperature dependence can be traced back
to a Schottky anomaly of $\alpha_i$ due to the thermal population
of the triplet state with $\Delta$ depending on uniaxial pressure
$p_i$. The opposite signs signal that $\partial \Delta/\partial
p_{010}>0$ and $\partial \Delta/\partial p_{10\overline{2}} <0$.
A magnetic field hardly changes $\alpha_i$ above about 10\,K, but
for lower $T$ well-defined anomalies develop for $H> 6$\,T, which
arise from the above-mentioned N\'{e}el order.
The anomalies are of opposite signs for both directions.
Fig.\,\ref{fig2} displays representative magnetostriction
measurements. The field-induced N\'{e}el order causes a kink in
$\epsilon_i$ leading to a jump-like change of $\partial
\epsilon_i/\partial H$ (again of opposite signs). The $\partial
\epsilon_i/\partial H$ curves obtained with increasing and
decreasing field hardly differ from each other and we do not
detect a hysteresis of $H_c$. Thus, our data agree with the
expected behavior of a second-order phase transition and disagree
with an ultrasound study showing a large hysteresis of order
0.5\,T for $H_c$.\cite{Sherman2003}

\begin{figure}[t]
\begin{center}
\includegraphics[width=0.32\textwidth]{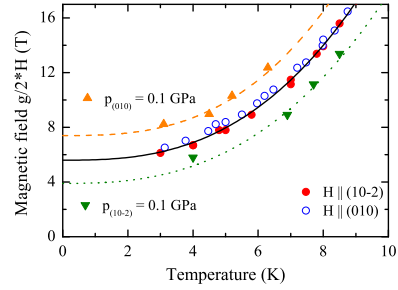}
\end{center}
\caption{Phase diagram of \tal\ from the magnetostriction and
thermal expansion measurements perpendicular to the $(010)$
($\circ$) and $(10\overline{2})$ ($\bullet$) planes. The triangles
display the calculated shift of $T_c$ and $H_c$ values for a
uniaxial pressure of 0.1\,GPa perpendicular to $(010)$
($\blacktriangle$) and $(10\overline{2})$ ($\blacktriangledown$).
The lines are power-law fits of the phase boundaries.}
\label{fig3}
\end{figure}

The phase diagram obtained from our data (Fig.\,\ref{fig3}) well
agrees with previous results.\cite{Oosawa2001} A power-law fit of
the form $(g/2)[H_c(T)-H_c(0)]\propto T^\Phi$ yields $\Phi=2.6$
and $H_c(0)=5.6$\,T (with the $g$ factor for the respective field
direction\,\cite{Oosawa2001}). The value of $\Phi$ agrees with
that obtained by Quantum Monte Carlo (QMC) simulations.
\cite{wessel2001nohadani2004} The QMC studies also reveal that
$\Phi $ sensitively depends on the temperature range of the fit,
what may explain why the experimental $\Phi$ obtained at finite
$T$ is much larger than the expected critical exponent $\Phi=1.5$
at $T \rightarrow 0$\,K for a Bose-Einstein condensation.

Using Ehrenfest relations our data together with specific heat
($C_p$) and magnetization ($M$) data allow to calculate the
uniaxial pressure dependencies
\begin{equation}
\frac{\partial T_c}{\partial
p_{i}}=V_{m}\,T_c\,\frac{\Delta\alpha_{i}}{\Delta C_p}\,\, \mbox{
and }\,\,
 \frac{\partial H_c}{\partial p_{i}}
 = V_{m}\frac{\Delta \frac{\partial \epsilon_i}{\partial
H}}{\Delta \frac{\partial M}{\partial H}}\,. \label{ehrenfest}
\end{equation}
Here, $V_m$ is the molar volume, $\Delta \alpha_{i}$ is the
anomaly size of $\alpha_i$ (see Fig.~\ref{fig1}) and ${\Delta
C_p}$ that of $C_p$\,\cite{Oosawa2001}, $\Delta \frac{\partial
\epsilon_i}{\partial H}$ is the slope change of $\epsilon_i$ at
$H_c$ (see Fig.~\ref{fig2}) and $\Delta \frac{\partial
M}{\partial H}$ that of $M$\,\cite{Shiramura1997}. In
Fig.\,\ref{fig3} we show some expected $T_c$ and $H_c$ values for
a hypothetical uniaxial pressure $p_i = 0.1$\,GPa with $p_i$
perpendicular to the $(010)$ ($\blacktriangle$) and
$(10\overline{2})$ ($\blacktriangledown$) planes. The N\'{e}el phase
is strongly destabilized by $p_{010}$ and stabilized by
$p_{10\overline{2}}$. The dashed lines represent power-law fits
of the pressure-shifted phase boundaries. Their extrapolations to
$T=0$\,K yield for the ambient-pressure spin gap
$\Delta=g\mu_BH/k_B\simeq 7.5$\,K and huge values for its
uniaxial pressure dependencies: $\partial \ln\Delta /\partial
p_{10\overline{2}}\simeq -180$\,\%/GPa and $\partial \ln\Delta
/\partial p_{010}\simeq +190$\,\%/GPa. The respective signs of
$\partial \ln\Delta /\partial p_i$ agree to those of the Schottky
anomalies of $\alpha_i$. Hydrostatic pressure also strongly
stabilizes the N\'{e}el phase,\cite{Oosawa2003atanaka2003a} but due
to the opposite signs the $\partial \ln\Delta /\partial p_i$
nearly cancel each other. Thus we conclude that uniaxial pressure
parallel to the $[201]$ direction, which is perpendicular to both
of our measurement directions, will also strongly decrease
$\Delta$. Analyzing the relation between the uniaxial pressure
dependencies of $\Delta$ and of the magnetic susceptibility
around 30\,K yields that
both arise from a pressure-dependent intradimer
coupling.\cite{niko}

In summary, the field- and temperature-dependent changes of
different lattice constants yield huge and strongly anisotropic
uniaxial pressure dependencies of the spin gap leading to drastic
changes of the phase diagram of \tal .

This work was supported by the DFG via SFB 608.



\begin{thebibliography}{99}




\bibitem{Nikuni2000matsumoto2002ruegg03}
T.~Nikuni et al, 
Phys.\ Rev.\ Lett. {\bf 84} (2000) 5868;
M.~Matsumoto et al, 
Phys.\ Rev.\ Lett. {\bf 89} (2002) 077203;
Ch. R\"{u}egg et al, 
Nature {\bf 423} (2003) 62.

\bibitem{tanaka2001Oosawa2002d}
H.~Tanaka et al, 
J.\ Phys.\ Soc.\ Japan {\bf 70} (2001) 939;
A.Oosawa et al, 
Phys.\ Rev.\ B {\bf 66} (2002) 104405.

\bibitem{Oosawa2003atanaka2003a}
A.~Oosawa et al, 
J.\ Phys.\ Soc.\ Japan {\bf 72}, (2003) 1026;
H.~Tanaka et al, 
Physica B {\bf 329} (2003) 697.

\bibitem{Sherman2003}
E.~Ya. Sherman et al, 
Phys.\ Rev.\ Lett. {\bf 91} (2003) 057201.

\bibitem{Oosawa2001}
A.~Oosawa et al, 
\newblock Phys.\ Rev.\ B {\bf 63}, 134416 (2001).

\bibitem{wessel2001nohadani2004}
S.~Wessel et al, 
 Phys.\ Rev.\ Lett. {\bf 87} (2001) 206407;
O.~Nohadani et al, 
Phys.\ Rev.\ B {\bf 69} (2004) 220402(R).

\bibitem{Shiramura1997}
W.~Shiramura et al, 
J.\ Phys.\ Soc.\ Japan {\bf 66} (1997) 1900.

\bibitem{niko}
N.~Johannsen et al, Phys. Rev. Lett. {\bf 95} (2005) 017205.






\end{thebibliography}


\end{document}